\documentclass[aps,prd,superscriptaddress,nofootinbib,amsmath,amsfonts,groupedaddress]{revtex4-2}

\usepackage{setspace}
\usepackage{graphicx}
\usepackage{subcaption}
\usepackage{epsfig}
\usepackage{bm}
\usepackage{amssymb}
\usepackage{float}
\usepackage{amsmath}
\usepackage{palatino}
\usepackage{dcolumn}
\usepackage[usenames,dvipsnames]{color}
\usepackage{enumitem}
\usepackage[mathscr]{eucal}
\usepackage{hyperref}
\hypersetup{colorlinks=true,linkcolor=blue,citecolor=red,urlcolor=blue}
\def\doi{http://doi.org}

\newcommand{\be}{\begin{equation}}
\newcommand{\ee}{\end{equation}}
\newcommand{\beano}{\begin{eqnarray*}}
\newcommand{\eeano}{\end{eqnarray*}}
\newcommand{\ba}{\begin{eqnarray}}
\newcommand{\ea}{\end{eqnarray}}

\newcommand{\om}{\ensuremath{\Omega_M}}

\def\chariteratehelpA#1 #2\relax{%
  \chariteratehelpB#1\relax\relax%
  \ifx\relax#2\else\rlap{\charop{~}}\ \chariteratehelpA#2\relax\fi
}
\def\chariteratehelpB#1#2\relax{%
  \charop{#1}%
  \ifx\relax#2\else
    \chariteratehelpB#2\relax%
  \fi
}
\def\charop#1{\def\stacktype{L}\def\useanchorwidth{T}%
  \stackon[0pt]{#1}{\scalebox{.85}[1]{\color{red}$\sim$}}}

\usepackage{xcolor, soul}
\sethlcolor{yellow}

\begin{document}

\title{ Stability analysis of a dark energy model in Rastall gravity}

\author{Shaily} 
\email{shaily.ma19@nsut.ac.in}
\affiliation{Department of Mathematics, Netaji Subhas University of Technology, New Delhi-110 078, India}
\affiliation{School of Computer Science Engineering and Technology, Bennett University, Greater Noida, India}
\author{Akanksha Singh} 
\email{akanksha.ma19@nsut.ac.in}
\affiliation{Department of Mathematics, Netaji Subhas University of Technology, New Delhi-110 078, India}
\author{J. K. Singh}
\email{jksingh@nsut.ac.in}
\affiliation{Department of Mathematics, Netaji Subhas University of Technology, New Delhi-110 078, India}
\author{Saddam Hussain}
\email{msaddam@iitk.ac.in}
\affiliation{Department of Physics, Indian Institute of Technology, Kanpur, Uttar Pradesh 208016, India}
\author{Ratbay Myrzakulov}
\email{rmyrzakulov@gmail.com}  
\affiliation{Ratbay Myrzakulov Eurasian International Centre for Theoretical Physics, Astana, Kazakhstan.}

\begin{abstract}
We study a cosmological model in Rastall's theory of gravity in the framework of the flat FLRW metric. We formulate the value of the Hubble parameter, which contains two model parameters, $ \alpha $ and $ j $. Employing the Markov Chain Monte Carlo (MCMC) sampling technique, we determine the values of these model parameters along with their uncertainties. Moreover, we derive the equation of state (EoS) parameter, which converges around the quintessence region. We perform a dynamical system analysis using the linearization technique to validate the results independently. Also, we discuss various physical attributes of the model, highlighting the transition to acceleration and the violation of the strong energy condition (SEC) in the late stages of evolution. In conclusion, our model mimics the behavior of a dark matter fluid during the past epoch and transitions into a quintessence dark energy model in the future epoch. 
	
\end{abstract}

\maketitle

\section{Introduction}\label{intro}
\qquad The General Theory of Relativity stands as a monumental achievement and is universally acknowledged as a cornerstone of contemporary physics. This theory, articulated in the framework of differential geometry, pioneered the integration of advanced mathematics into physical theories, setting the stage for subsequent gauge and string theories \cite{Wheeler:1990nd, Geroch:1978bu}. General relativity has redefined the expectations of a physical theory and has profoundly transformed our comprehension of the cosmos. For over a hundred years, it has stood as the absolute theory governing gravitational interactions, aligning with all experimental data and accounting for a vast array of observational findings \cite{Dyson:1920cwa, Will:2014zpa}. However, certain theoretical and observational challenges persist, such as the cosmic acceleration and the enigmatic presence of dark matter and energy, necessitating the consideration of potential modifications of general relativity \cite{Einstein:1916, Debono:2016vkp}.

Various modifications and alternatives to the theory of gravity have been proposed, typically arising from two distinct methodologies. The first approach retains the fundamental postulates of general relativity while augmenting the Lagrangian density with additional terms, resulting in modified field equations \cite{Singh:2022eun, Singh:2023gxd, Singh:2022jue, Singh:2018xjv, Singh:2022gln, Singh:2019fpr, Singh:2022wwa, Nagpal:2018mpv, Nojiri:2010wj, Nojiri:2006ri, Nojiri:2017ncd, deHaro:2023lbq, Bamba:2012cp}. The second approach involves altering the fundamental assumptions of general relativity. Rastall gravity is an example of this latter category. Unlike the conventional perspective, which assumes a null covariant divergence of the energy-momentum tensor (EMT), Rastall proposed a different perspective, arguing that the zero divergence of the Einstein tensor does not necessarily lead to the zero divergence of the EMT \cite{Rastall:1972swe}. This implies a departure from the principle of energy conservation. However, this relaxation of the zero divergence condition does not undermine the successes of general relativity. Rastall gravity sought to revise the theory of general relativity by positing that the covariant derivative of the EMT is proportional to the derivative of the Ricci scalar \cite{Rastall:1972swe}, thereby explicitly introducing a non-minimal coupling between matter and geometry in curved spacetimes.

A significant drawback of Rastall gravity lies in its absence of a Lagrangian formulation. However, some studies in the literature suggest a potential solution by formulating a Lagrangian-based model akin to Rastall theory as an MG model where the Lagrangian is dependent on the Ricci scalar $ R $ and the trace of the energy-momentum tensor $ T $. There are ongoing debates about whether Rastall gravity is a genuine modification of general relativity or merely a redefinition of the energy-momentum tensor \cite{Lindblom:1982, Visser:2017gpz}. Geometrically, Rastall gravity aligns with Einstein's gravity, unimodular theory, and $ f(R,T) $ theory, but it’s been repeatedly emphasized that the physical implications are significantly different \cite{Darabi:2017coc}.

Dynamical system analysis serves as a crucial tool in cosmological studies, offering a qualitative method to explore the behavior of solutions to cosmic models without drowning in complex calculations. Often, this method is employed to determine both the stability of the model and the existence of critical points \cite{SantosDaCosta:2018bbw, Olmo:2005hc}. By examining the asymptotic behavior of critical points of the dynamical system, one can explain the overarching dynamics of the universe across different cosmological epochs \cite{Duchaniya:2022rqu, Kofinas:2014aka, Kadam:2022lgq}. Researchers like Kadam et al. have presented an exploration of critical points representing different cosmic eras, including the matter-dominated, radiation-dominated, and dark energy eras in $ f(T,B) $ gravity \cite{Kadam:2023ufk}. Stachowski and Szyd\l{}owski have tackled the dynamics of models by running cosmological constant terms with the help of a dynamical analysis approach \cite{Stachowski:2016dfi}. Roy and Banerjee's \cite{Roy:2017mnz} extensive analysis of the stability of cosmological solutions for a spatially flat homogeneous and isotropic cosmological model within an extended Brans-Dicke theory through a phase space analysis serves as yet another testament to the power of dynamical system methods. Moreover, the versatility of dynamical systems extends to exploring interacting dark energy scenarios, offering a unified framework for understanding the cosmos at both background and perturbation levels \cite{Khyllep:2022spx, Khyllep:2021wjd, Landim:2019lvl, Alho:2019jho}. 

The dynamical systems method is applied to probe critical points, and these points offer insights into the transitions between accelerating and decelerating epochs, characterizing phases dominated by dark energy, radiation, and matter \cite{Singh:2022ocv}. Recent studies have promoted the discussion about the stability of various cosmological models using dynamical system analysis. For instance, Roy and Banerjee \cite{Roy:2014hsa} scrutinized the stability criteria of chameleon models within the context of spatially homogeneous and isotropic cosmology, employing the tools of autonomous system dynamical analysis. Similarly, Das et al. \cite{Das:2023rat} explored the stability of the warm quintessential dark energy model. The collection of recent studies in modified gravity theories underscores the pivotal role of dynamical system analysis in advancing our understanding of the cosmos \cite{Chakraborty:2018bxh, Odintsov:2017tbc, Alho:2016gzi, Carloni:2015jla, Odintsov:2018uaw, Oikonomou:2019muq, Roy:2014wqa, Bahamonde:2017ize, Shabani:2017vns, Shabani:2023xfn, Shabani:2023nvm, Oikonomou:2017ppp, Chatzarakis:2019fbn, Duchaniya:2023aeu, Lohakare:2023ocb, Kadam:2023bpb, Pati:2022dwl, Narawade:2023odv, Bhagat:2023kvf, De-Santiago:2012ibi, Dutta:2018xcz, Carloni:2015lsa, Boehmer:2015sha, Chatterjee:2021ijw, Hussain:2023kwk, Mirza:2014nfa}.

Moreover, the intriguing theory of Rastall gravity has also been subjected to dynamical system analysis \cite{Shabani:2022lxd, Silva:2012gn, Singh:2021liv, Shabani:2022zlx, Singh:2022ocv, Khyllep:2019odd}. Shabani et al. utilize the dynamical system approach to conduct a comprehensive exploration of the universe's evolution under the assumption of a particular form for the Rastall parameter $\lambda$ \cite{Shabani:2022zlx}. In exploring the Rastall model within non-flat Friedmann-Robertson-Walker (FRW) spacetime, incorporating a barotropic fluid, researchers delve into an autonomous system analysis. The literary work by Khyllep and Dutta \cite{Khyllep:2019odd} attempts to contribute to the ongoing debate surrounding the equivalence between Rastall gravity and general relativity. This is achieved through an in-depth analysis of the evolution of the Rastall-based cosmological model, both at the background and perturbation levels, by applying dynamical system techniques.

Over the past decades, Rastall gravity has attracted attention from researchers, resulting in a plethora of scholarly articles. Moreover, it includes a parameter that measures its deviations from general relativity. In the past ten years, there has been a notable surge in interest surrounding the exploration of observational constraints on Rastall's parameter, along with the study of applications across various cosmological contexts, including black holes and other compact celestial bodies. The characteristics of the black hole solutions in Rastall gravity can be found in the literature \cite{Ghosh:2021byh, Sakti:2019krw, Sakti:2021gru, Guo:2021bwr, Nashed:2022dkj, Gogoi:2021dkr, Shao:2022oqv, Narzilloev:2022bbs}. Rastall gravity offers some interesting results; for instance, de Sitter black holes have been found without explicitly assuming a cosmological constant \cite{Heydarzade:2016zof}. It has been demonstrated that Rastall gravity accommodates the accelerated expansion of the universe. Several recently published literary works focus on different aspects of this theory \cite{Singh:2020akk, Singh:2021liv, Singh:2022ocv, Singh:2022wvz}.

In this paper, we assemble our work in the following manner: Sec. \ref{RG} comprises the brief background of Rastall gravity theory. In Sec. \ref{FormulationHz}, the formulation for the Hubble parameter in terms of the jerk parameter takes place. Afterward, in Sec. \ref{sectionEFE}, we obtain the Einstein field equations (EFEs) for the flat FLRW space-time metric in Rastall gravity theory. Observational analysis has been done in Sec. \ref{sectionObs}, where three observational datasets are used to obtain the best-fitted values of model parameters. Sec.  \ref{sectionplots} comes along with the investigation of the physical and dynamical behavior of the universe. In Sec. \ref{dss}, we accomplish a meticulous dynamical system analysis to discover the stability of the system depending on the various model parameters. In Sec. \ref{con}, we conclude our findings. Finally, in the Appendix. \ref{ap}, we consider the autonomous equation in 1D and determine the stability of the system around the fixed point.

\section{ Rastall Gravity}\label{RG}
In this section, we aim to obtain a solution while taking into consideration the physical background of the gravity sector together with the matter source. We include two parts of the model: the Rastall field equations and the matter field equations, to start the computation in the Rastall gravity sector. In this modification of General Relativity (GR), the covariant conservation equation of the energy-momentum tensor expressed as $ \nabla_j T^{ij} = 0 $ was changed to a more generalized version as \cite{Sekhmani:2024jli}
\begin{equation}\label{1}
\nabla_j T^{ij} = u^i.
\end{equation}
To make the theory consistent with GR, the right-hand side of the Eq. (\ref{1}) must be equal to zero when the scalar curvature or the background curvature vanishes. Thus, for our convenience, we can take the four-vector $u^i$ as,
\begin{equation}\label{2}
u^i=\lambda \nabla_i R,
\end{equation}
where $ \lambda $ is a free parameter known as Rastall’s parameter. From Eqs. (\ref{1}) and (\ref{2}), we have the Rastall field
equation as
\begin{equation}\label{3}
R_{ij}-\frac{1}{2}\left(1-2\beta \right)Rg_{ij} = \kappa T_{ij},  
\end{equation}

where $ \beta = \kappa \lambda $ and $ \kappa $ is the gravitational coupling constant. The trace of the Eq. (\ref{3}) yields
\begin{equation}\label{4}
R=\frac{\kappa}{4\beta-1}T,~\beta \neq 1/4.
\end{equation}

The field equation in the presence of a non-vanishing cosmological constant $ \Lambda $ is
\begin{equation}\label{5}
E_{ij} + \Lambda g_{ij} + \beta g_{ij} R = \kappa T_{ij},
\end{equation}
where $ E_{ij} $ is the standard Einstein tensor.


\section{ Formulation for the Hubble parameter in terms of the jerk parameter}\label{FormulationHz}

In physics, the term jerk, also referred to as jolt, denotes the rate at which an object’s acceleration changes over time. It is a vector quantity and is symbolized by $ j $. Its units are $ m/s^3 $ in the International System of Units (SI) or $ g_0/s $ in terms of standard gravities per second. As a vector, jerk $ j $ is mathematically defined as the first derivative of acceleration, the second derivative of velocity, and the third derivative of position with respect to time:

\begin{equation}\label{6}
j(t) = \frac{\mathrm{d} \mathbf q(t)}{\mathrm{d}t} = \frac{\mathrm{d}^2 \mathbf v(t)}{\mathrm{d}t^2} = \frac{\mathrm{d}^3 \mathbf r(t)}{\mathrm{d}t^3},
\end{equation}
where $ q $, $ v $, $ r $, and $ t $ represent acceleration, velocity, position, and time, respectively.

Differential equations of the third order,
\begin{equation}\label{7}
J\left(\overset{\mathbf{...}}{x}, \ddot{x}, \dot{x}, x\right) = 0,
\end{equation}
are colloquially termed jerk equations. When transformed into a system of three ordinary first-order non-linear differential equations, jerk equations represent the fundamental framework for modeling chaotic phenomena. This characteristic piques mathematical curiosity in jerk systems. Correspondingly, systems involving derivatives of the fourth order or higher are classified as hyper-jerk systems \cite{Chlouverakis:2006}.

If we look at the jerk parameter, denoted by $ j $, as a function of the redshift $ z $, there exists a specific relationship between the jerk parameter $ j $ and the deceleration parameter $ q $, which is given by
\begin{equation}\label{8}
j(z) = q(z)+2q(z)^2+(1+z)\frac{dq(z)}{dz}.
\end{equation}
When the jerk parameter is constant and equals $ 1 $, it corresponds to a flat $ \Lambda $CDM model. Without loss of generality, we consider the jerk parameter $ j $ as an arbitrary constant \cite{AlMamon:2018uby, Myrzakulov:2023ohc} and solving the differential equation (\ref{8}), we get 
\begin{equation}\label{9}
\frac{\sqrt{8j+1}+4q+1}{\sqrt{8j+1}-4q-1} = \left(\beta+\beta z\right)^{\sqrt{8j+1}},
\end{equation}
where $ \beta $ is an integration constant.

Rewriting Eq. (\ref{9}) and using $ \beta = 1, $ we get
\begin{equation}\label{10}
    q = \frac{-1}{4} + \frac{\sqrt{8j+1}}{4} - \frac{\sqrt{8j+1}}{2\left(1+\left(1+z\right)^{\sqrt{8j+1}}\right)}.
\end{equation}

The equation representing the deceleration parameter $ q $ in terms of the Hubble parameter $ H $ and redshift $ z $ is
\begin{equation}\label{11}
	q(z) = -1 +\frac{(1+z)}{H(z)}\frac{dH}{dz}.
\end{equation}

Using Eq. (\ref{10}) in Eq. (\ref{11}), we get
\begin{equation}\label{12}
	\frac{1}{1+z}\left[ 1 - \frac{1}{4} + \frac{\sqrt{8j+1}}{4} - \frac{\sqrt{8j+1}}{2\left(1+\left(1+z\right)^{\sqrt{8j+1}}\right)} \right] dz = \frac{dH}{H}.
\end{equation}

By performing integration on both sides of the Eq. (\ref{12}), we obtain
\begin{equation}\label{13}
H = \alpha \left(1+z\right)^{\frac{3+\sqrt{8j+1}}{4}} \sqrt{1+\left(1+z\right)^{-\sqrt{8j+1}}},
\end{equation}
where $ \alpha $ is an integration constant.


\section{Formulation of Einstein Field Equations in Rastall Gravity}\label{sectionEFE}

\qquad Rastall introduced the fact that the evolution of the universe does not perturb the rate of energy-momentum transference between matter fields and geometry. In this regard, Rastall supposed a coupling coefficient between the divergence of EMT and the derivative of the Ricci scalar, which is a constant and known as the Rastall parameter. This theory also describes the dark energy section to explain the accelerated universe. From Eq. (\ref{5}), the generalized field equations for Rastall gravity theory in the absence of the cosmological constant $ \Lambda $ are given by
\begin{equation}\label{14}
R^j_i-\frac{1}{2}\delta_{ij} R=\kappa (T^j_i-\lambda \delta_{ij} R),
\end{equation}
or
\begin{equation}\label{15}
R_{ij}-\frac{1}{2}\left(1-2\kappa \lambda\right)Rg_{ij} = \kappa T_{ij},
\end{equation}

where $ \lambda $ is the Rastall parameter and $ \kappa $ is the gravitational coupling constant. For flat isotropic Universe, the FLRW metric is written as
\begin{equation}\label{16}  
ds^2=-dt^2+a^2 (t)(dx^2+dy^2+dz^2),
\end{equation}
where $ a(t) $ stands for the scale factor of the universe. For the specific values of $ \kappa $ and  $ \lambda $, we can come back to the theory of general relativity. Now, since Hubble parameter $ H=\dot{a}/a $, therefore using Eqs. (\ref{15}), (\ref{16}), we get the Einstein field equations for Rastall gravity theory in terms of $ H $ as

\begin{equation}\label{17}
3(1-4\kappa \lambda)H^2+2(1-3\kappa \lambda)\dot{H} = -\kappa p,
\end{equation}
\begin{equation}\label{18}
3(1-4\kappa \lambda)H^2-6 \kappa \lambda\dot{H} = \kappa \rho,
\end{equation}
where $ \rho $ is the energy density, $ p $ is the isotropic pressure, and $ \dot{H} $ indicates the derivative of $ H $ with respect to cosmic time $ t $.

Using the relation $ \dot{H} = -(1+z)H(z)\frac{dH}{dz} $ and Eq. (\ref{13}) in Eqs. (\ref{17}), (\ref{18}) we get
\begin{equation}\label{19}
\rho(z) = \frac{3\left(1+z\right)^{\frac{3-\sqrt{1+8j}}{2}}\alpha^2 \left[\left(2-5\kappa\lambda\right)\left(1+\left(1+z\right)^{\sqrt{1+8j}}\right) - \kappa\lambda\sqrt{1+8j}\left(1-\left(1+z\right)^{\sqrt{1+8j}}\right)\right]}{2\kappa}
\end{equation}
and
\begin{equation}\label{20}
p(z) = \frac{-\left(1+z\right)^{\frac{3-\sqrt{1+8j}}{2}}\alpha^2 \left[ 3\left(1-5\kappa\lambda\right) \left(1+\left(1+z\right)^{\sqrt{1+8j}}\right) + \sqrt{1+8j}\left(1-3\kappa\lambda\right) \left(1-\left(1+z\right)^{\sqrt{1+8j}}\right) \right)}{2\kappa},
\end{equation}
and the EoS parameter is given by $ \omega(z)= p(z)/\rho(z) $, whose variations with respect to redshift $ z $ can be seen in Fig. \ref{rhopw}(c).

\section{Observational Analysis}\label{sectionObs} 
\qquad Observational analysis is one of the advantageous parts of cosmology, enabling us to determine the best-fitted results for the parameters of the model. In this analysis, we use the standard datasets to compute the optimal solution of the model parameters. Here, in this paper, we are using the Hubble dataset $ H(z) $ (with $ 77 $ points), the $ Pantheon $ dataset (with $ 1048 $ points), and their joint dataset ($ H(z)+Pantheon $), to calculate the best-fit value of $ \alpha $ and $ j $. For this purpose, we use the Markov Chain Monte Carlo (MCMC) method, and the coding to plot the contour has been done in Python with the EMCEE library.

\subsection{$ H(z) $ Dataset}
\qquad $ H(z) $ data acquired from cosmic chronometers, which play an important role in studying the dark division of the universe \cite{Chimento:2007da}. In literature, various models have been discussed with different numbers of $ H(z) $ data points \cite{Singh:2022nfm, Shaily:2024nmy, Singh:2024aml, Singh:2024kez, Balhara:2023mgj}. The presented model uses $ H(z) $ dataset having $ 77 $ data points \cite{Shaily:2022enj}. We use the $ \chi^2 $ test to obtain the best-fit value of the model parameters, and the formula for $ \chi^2 $ is written as
\begin{equation}\label{21}
\chi _{Hub}^{2}(j,\alpha)=\sum\limits_{i=1}^{77} \frac{[H(j,\alpha,z_{i})-H_{obs}(z_{i})]^2}{\sigma _{z_i}^2},
\end{equation}

where $ H(j,\alpha,z_{i}) $ and $ H_{obs} $ depict the theoretical and observed values of the Hubble parameter, respectively, and $ \sigma_{(z_{i})} $ shows the standard deviation for every $ H(z_i) $.

The analysis of the function $ \chi_{Hub}^{2} $ using the Hubble dataset has been found to be not very reliable for any cosmological models. Hence, we opt for more reliable observation using the $ Pantheon $ and their joint datasets.
\begin{table}[H]
\caption{Model Parameters: Best-Fit Results}
\begin{center}
\label{tabparm}
\begin{tabular}{l c c c r} 
\hline\hline
\\ 
{Dataset} &      ~~~~~  $ j $ & ~~~~~ $ \alpha $  
\\
\\
\hline      
\\
{$ H(z) $ }     & ~~~~~ $ 0.437 \pm 0.046 $   &  ~~~~~ $ 46.24 \pm 0.43 $ 
\\
\\
{$ Pantheon $ }     &  ~~~~~ $ 0.95 \pm 0.000094 $   &  ~~~~~ $ 50.000008^{+0.000076}_{-0.000097} $ 
\\
\\
{ $H(z)+Pantheon $}  & ~~~~~ $ 1.09999^{+0.00011}_{-0.000098} $   &  ~~~~~ $ 51.000001^{+0.000097}_{-0.000085} $
\\
\\ 
\hline\hline  
\end{tabular}    
\end{center}
\end{table}

\subsection{Pantheon Dataset}
\qquad From the beginning of the $ 21^{st} $ century, Type Ia supernovae have been very useful for discussing the cosmological parameters, as this is the first candidate to evidence the accelerated expansion of the universe. Supernovae are stars that, upon explosion, emit a vast amount of energy and expand their outer shell. This aspect is related to the study of the time evolution of their spectrum and brightness. In our model, we use the $ Pantheon $ dataset ($ 1048 $ data points) and constrain the value of model parameters \cite{Pan-STARRS1:2017jku}. The $ Pantheon $ dataset is the collection of data that have been compiled by different surveys, e.g., the CfA1-CfA4 surveys, the Supernovae Legacy Survey (SNLS), the Pan-STARRS1 (PS1), the Carnegie Supernova Project (CSP), the Sloan Digital Sky Survey (SDSS), various Hubble Space Telescope (HST) \textit{etc.} \cite{Riess:1998dv, Jha:2005jg, Hicken:2009df, Contreras:2009nt, SDSS:2014irn}. The range of redshift is between $ 0.01 $ and $ 2.26 $ in these surveys. In the analysis of Type Ia supernovae, we deal with both redshift and luminosity distance, which are closely related to standard candles in the universe. The luminosity distance is
\begin{equation}\label{22}
D_L(z)=(1+z) \int_0^z \frac{c}{H(z^*)}dz^* ,
\end{equation}

where $ c $ is the speed of light. And the apparent magnitude can be written in terms of luminosity distance as

\begin{equation}\label{23}
    m(z)=M+5log_{10}\Bigg[\frac{D_L(z)}{1Mpc}\Bigg]+25,
\end{equation}
where $ M $ is the absolute magnitude. Here, a dissipation between $ H_0 $ and $ M $ can be observed, and since the distance modulus $ \mu =m-M $, the formula for $ \chi^2 $ is written as
\begin{equation}\label{24}
\chi_{Pan}^{2}(j,\alpha)=\sum\limits_{i=1}^{1048}\left[ \frac{\mu_{th}(j,\alpha,z_{i})-\mu_{obs}(z_{i})}{\sigma _{\mu(z_{i})}}\right] ^2,
\end{equation}
where $ \mu_{th} $ and $ \mu_{obs} $ are used for theoretical and observed distance moduli, respectively. 

\begin{figure}\centering
	\subfloat[]{\label{a}\includegraphics[scale=0.75]{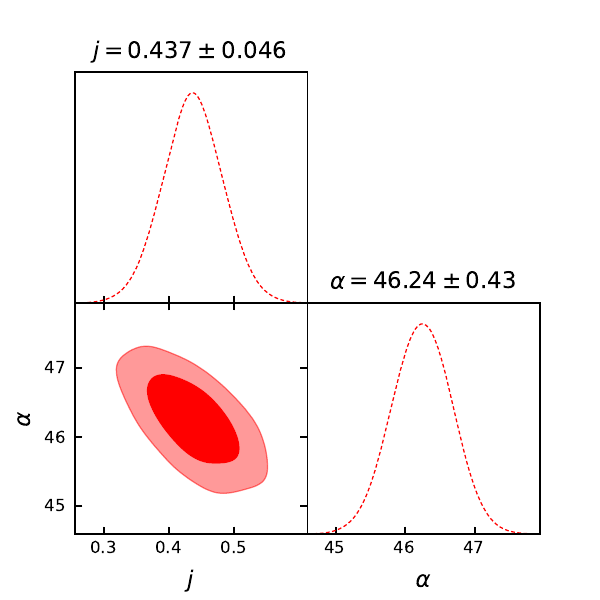}}\hfill
	\subfloat[]{\label{b}\includegraphics[scale=0.75]{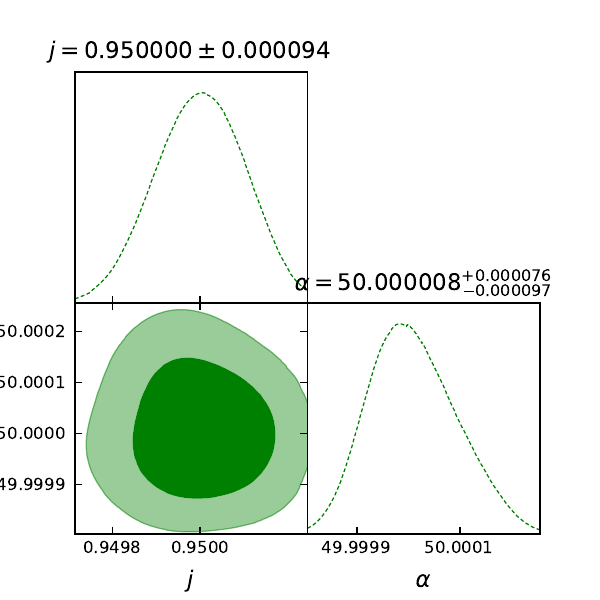}}\par
	\subfloat[]{\label{c}\includegraphics[scale=0.75]{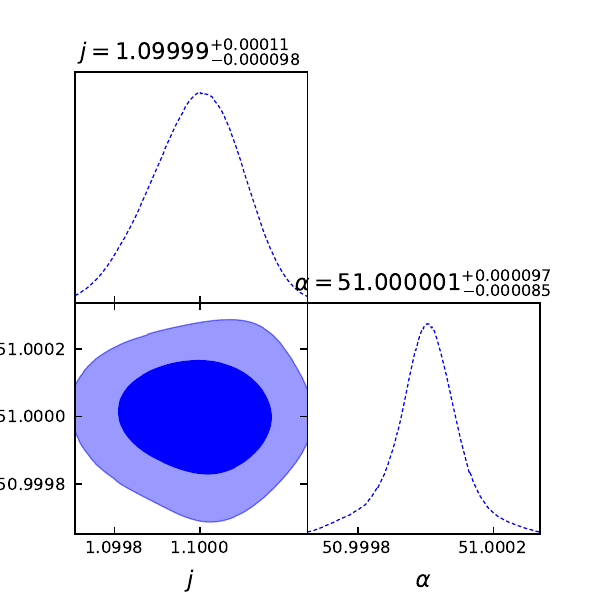}}\hfill
\caption{ Depiction of likelihood contours for the $ H(z) $, $ Pantheon $, and $ H(z)+Pantheon $ datasets within the $ 1\sigma $ and $ 2\sigma $ confidence levels.}
\label{contours}
\end{figure}

\subsection{Joint Dataset}
To obtain better constraint values of the model parameters, we apply a combined dataset. The results are computed using the $ \chi_{HP}^2 $ function, which is the sum of the $ \chi_{Hub}^{2} $ and $ \chi_{Pan}^{2} $ functions, as shown below
\begin{equation}\label{25}
\chi _{HP}^{2}(j,\alpha)=\chi _{Hub}^{2}(j,\alpha)+\chi _{Pan}^{2}(j,\alpha).
\end{equation} 

The tabulated results in Table-\ref{tabparm} showcase the acquired values of $ j $ and $ \alpha $ based on the aforementioned observational datasets. While the $ \Lambda $CDM model has the present value of the jerk parameter $ j $ as $ 1 $, our model obtains the best-fit value of $ j $ close to $1$ for the $ Pantheon $ and joint datasets, and for the $ H(z) $ dataset, it is approximately $ 0.437 $.
 
\begin{figure}\centering
	\subfloat[]{\label{a}\includegraphics[scale=0.40]{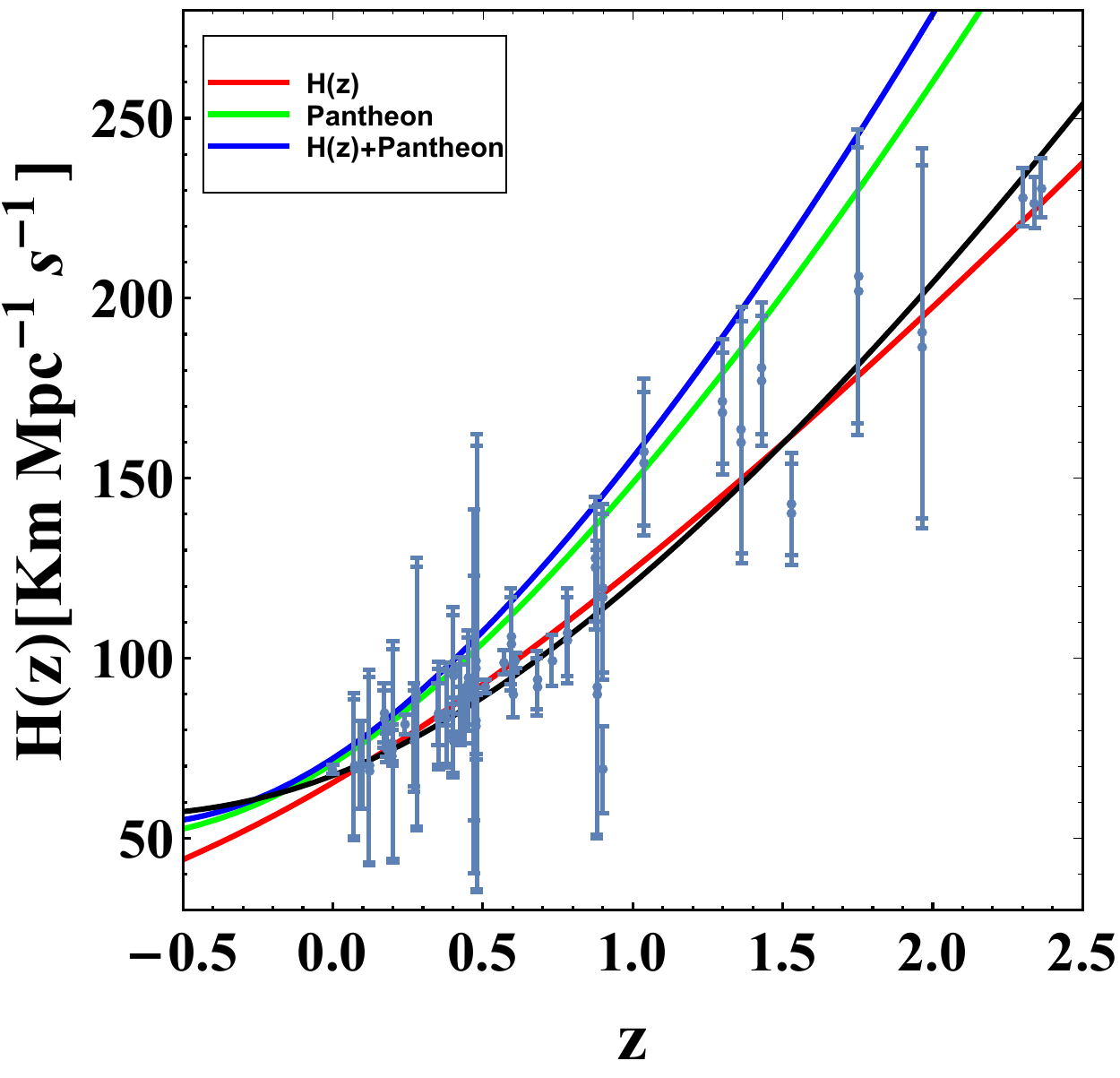}}\hfill
	\subfloat[]{\label{b}\includegraphics[scale=0.45]{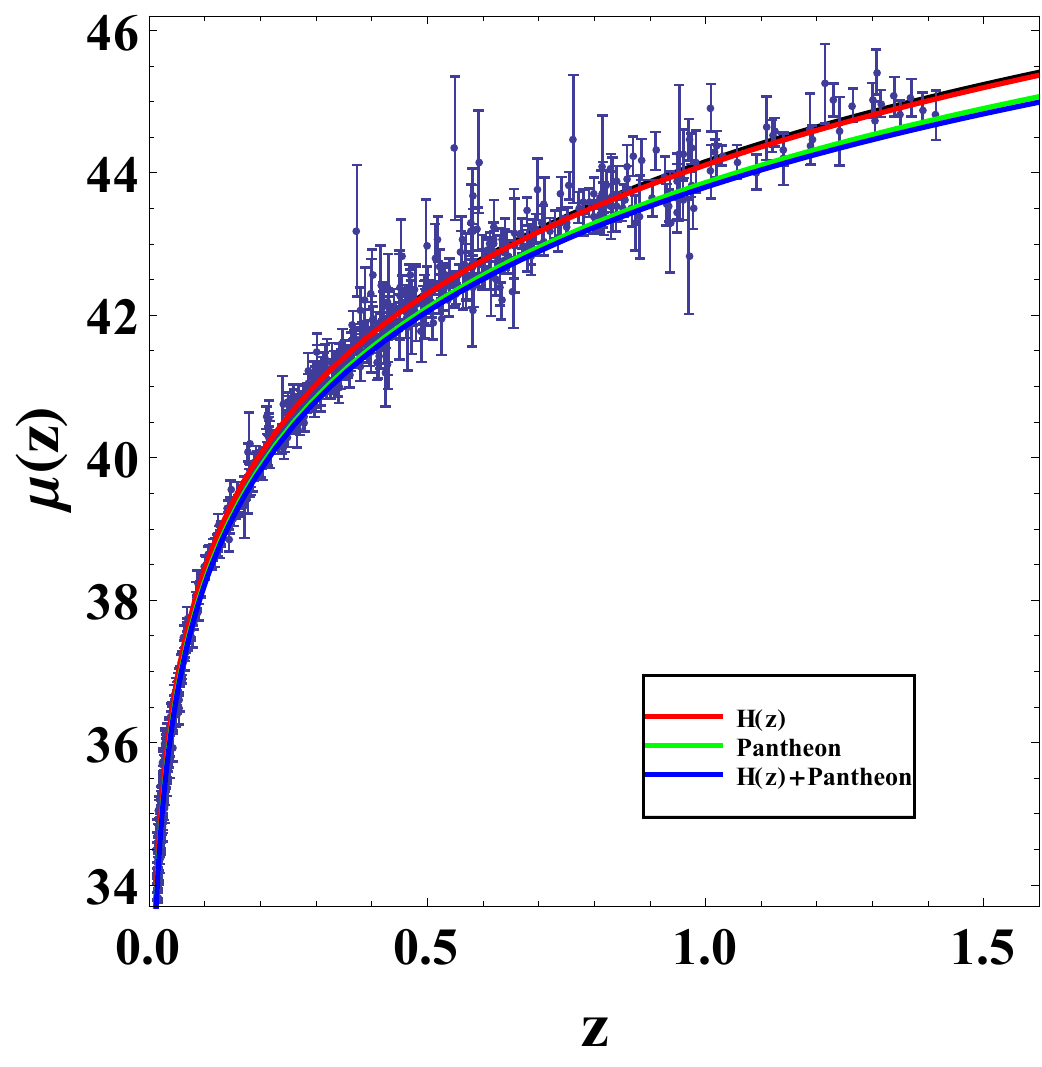}} 
\caption{ Visual display of error bars for $ H(z) $ and Type Ia Supernovae datasets.}
\label{errorbars}
\end{figure}

As the $ \Lambda $CDM model is one of the standard models in cosmology, by comparing our model with the $ \Lambda $CDM model, we check the consistency of our model, and for this purpose, error bar plots play a significant role. In our model, we have drawn the error bar plots for Hubble datasets and Type Ia Supernovae datasets. In Fig. \ref{errorbars}(a) and Fig. \ref{errorbars}(b), we observe the similarity between our model and the $ \Lambda $CDM model.

\section{Physical and Dynamical Components of the model}\label{sectionplots}
 
\begin{figure}[t]\centering
	\includegraphics[scale=0.46]{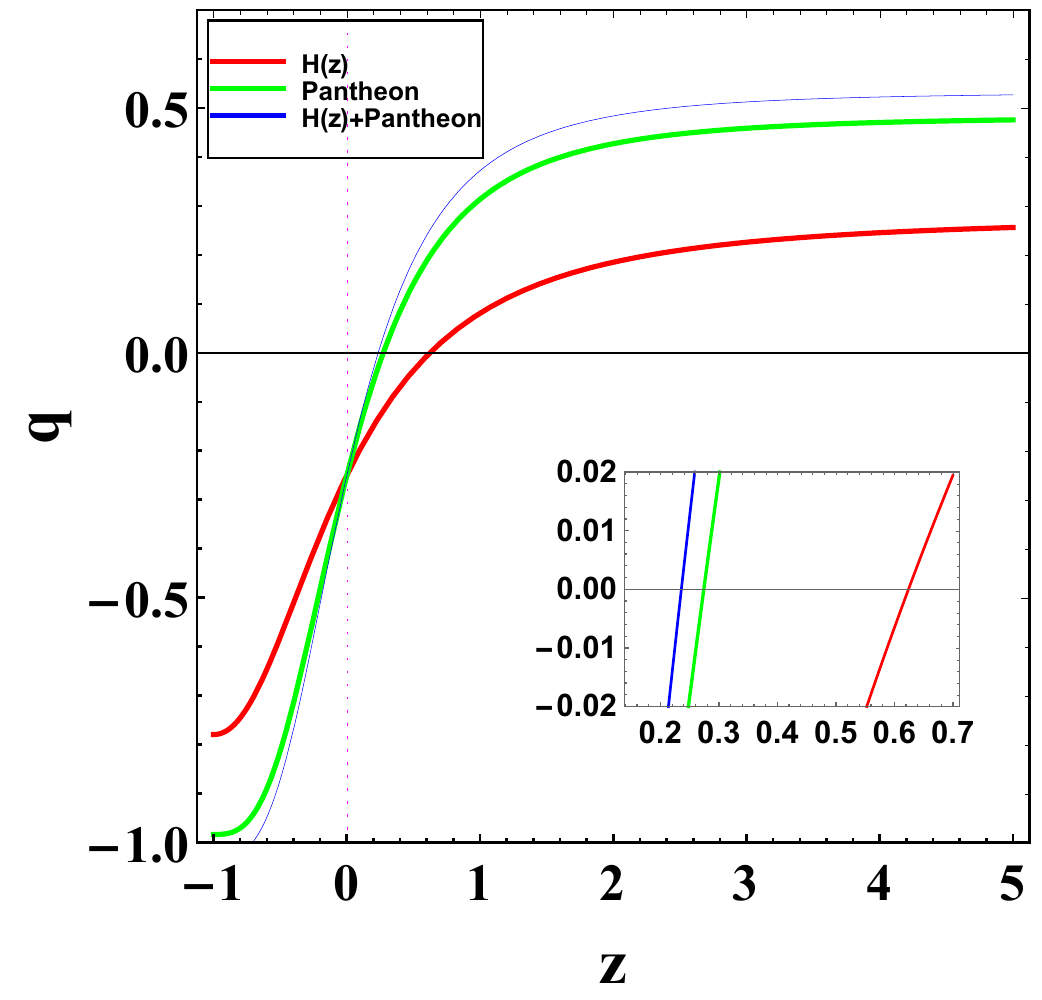}\hfill
\caption{ Plot of $ q $ \textit{\ vs.} $ z $ for the mentioned observational datasets.}
\label{dec}
\end{figure}

\qquad To explore the dynamical progression of the universe, let us start with one of the essential parameters named as the deceleration parameter ($ q $). The deceleration parameter has been calculated in Eq. (\ref{10}), and now we plot the curves for the obtained best-fit values of $ j $ and $ \alpha $. For all three mentioned observational datasets, we perceive that $ q $ is behaving alike, \textit{i.e.}, in Fig. \ref{dec}, at early times each trajectory is in the deceleration phase, and in late times as well as at present, our model shows the accelerating phase of the universe. At present, the value of $ q $ is $ -0.25 $  for $ H(z) $, $ Pantheon $, and joint datasets, and the value of redshift at the time of phase transition is $ 0.628 $ for $ H(z) $ dataset, $ 0.277 $ for $ Pantheon $ dataset, and $ 0.239 $ for $ H(z) + Pantheon $ dataset, which is clearly noticeable in the subplot of Fig. \ref{dec}.

It is widely accepted that following the Big Bang, the evolution of the universe is marked by two significant epochs: the radiation-dominated era and the matter-dominated era. The radiation-dominated era is required to anticipate primordial nucleosynthesis, whereas in the matter-dominated era, element creation takes place. Therefore, keeping these points in mind, we analyze the evolution of diverse physical parameters within our model.

\begin{figure}[t]\centering
	\subfloat[]{\label{a}\includegraphics[scale=0.45]{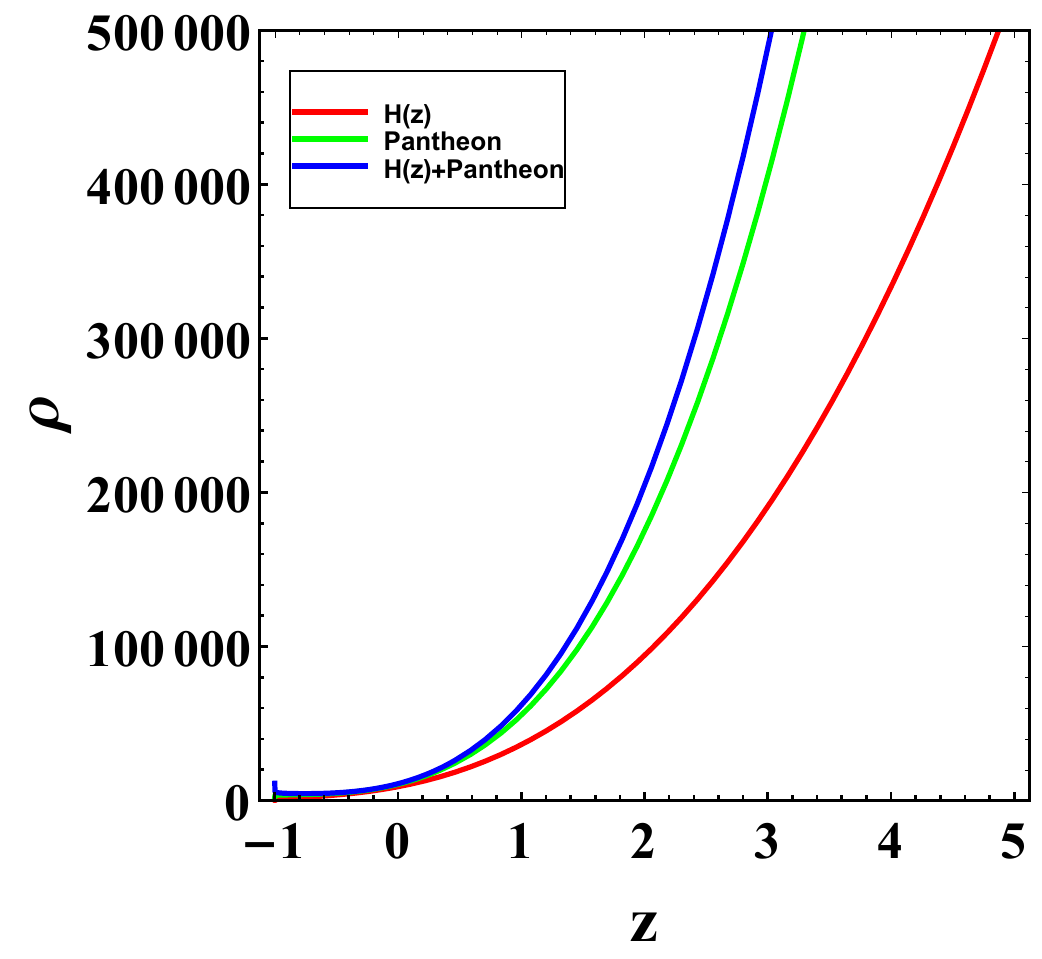}}	\hfill
	\subfloat[]{\label{b}\includegraphics[scale=0.47]{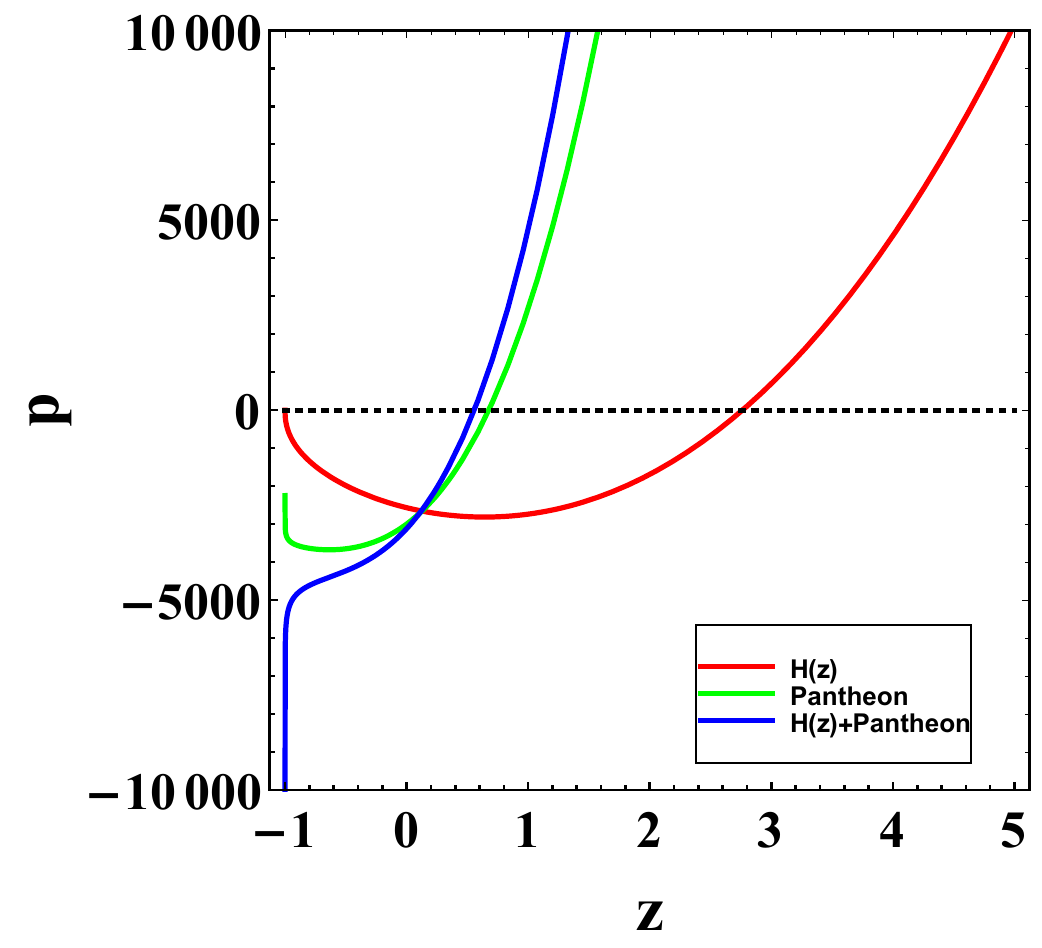}}\par
	\subfloat[]{\includegraphics[scale=0.45]{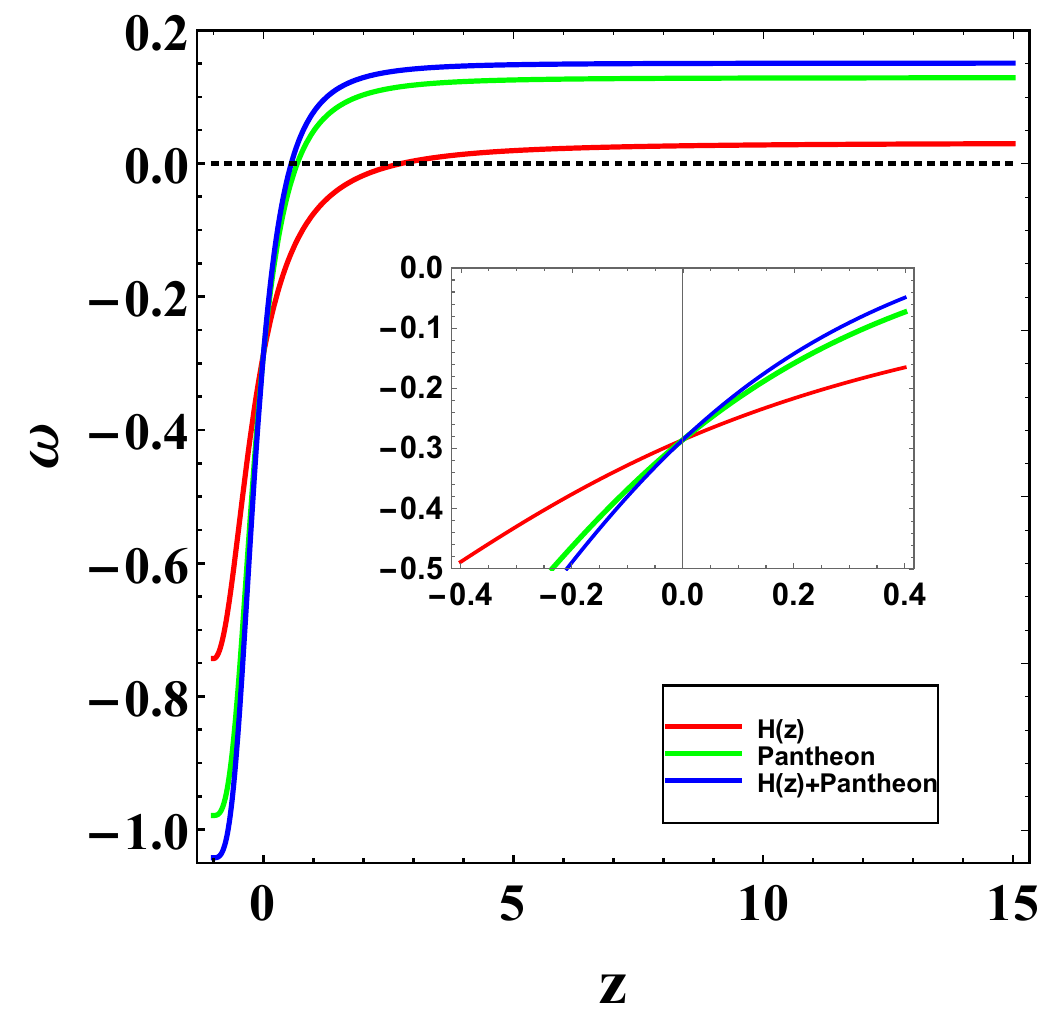}}
\caption{Visual display of $ \rho $, $ p $, and $ \omega $ \textit{\ vs.}  $ z $ for the mentioned observational datasets.}
\label{rhopw}
\end{figure}

For the fixed values $ \kappa=1 $ and $ \lambda=0.12 $, the energy density and isotropic pressure have been plotted using Eqs. (\ref{19}) and (\ref{20}) in Fig. \ref{rhopw} for the obtained best-fit values of model parameters. In Fig. \ref{rhopw}(a), we depict the change in energy density across varying redshifts. As expected, the energy density decreases from early to late times for all datasets. In this model, the isotropic pressure is plotted in Fig. \ref{rhopw}(b), which highlights that in the early universe, the value of pressure is highly positive, and at present as well as in the future, pressure is negative for all datasets. According to standard cosmology, the negative pressure reveals the expanding acceleration of the cosmos. The different value of the EoS parameter ($ \omega $) indicates the different forms of the universe, for example, $ \omega=0 $ $ \to $ pressureless matter, $ \omega\in (0, \frac{1}{3}) $ $ \to $ hot matter, $ \omega=\frac{1}{3} $ $ \to $ radiation, $ \omega\in (\frac{1}{3},1) $ $ \to $ hard universe, $ \omega=1 $ $ \to $ stiff matter, $ \omega >1 $ $ \to $ ekpyrotic matter, $ \omega\in (-1,-\frac{1}{3}) $ $ \to $ quintessence universe, $ \omega=-1 $ $ \to $ cosmological constant, and $ \omega<-1 $ $ \to $ phantom universe. In our model, one can observe the different forms of the universe with respect to the changes in the value of $ \omega $ (see Fig. \ref{rhopw}(c)). Today and in the future, our model aligns with a quintessence dark energy model.

In addition to the fascinating fields of study, there exist several major points for exploration, including the characterization of singularities, causal structure, and energy conditions. Originating from the renowned Raychaudhuri equation, energy conditions play a pivotal role in explaining the contribution of further geometrical terms in the stress-energy tensor, such as effective pressure and effective energy density. We study four energy conditions, which are the diminution of time-like or null vector fields concerning the energy-momentum tensor and the Einstein tensor obtained from Einstein field equations. Generally, four energy conditions are discussed in cosmological models namely, Weak energy condition (WEC), Null energy condition (NEC), Strong energy condition (SEC), and Dominant energy condition (DEC). These ECs are defined as: WEC: $ \rho \geq 0,~ \rho+p > 0 $, NEC: $ \rho+p \geq 0 $, SEC: $ \rho+3 p \geq 0 $, and DEC: $ \rho \geq |p| $.

\begin{figure}[t]\centering
	\subfloat[]{\label{a}\includegraphics[scale=0.43]{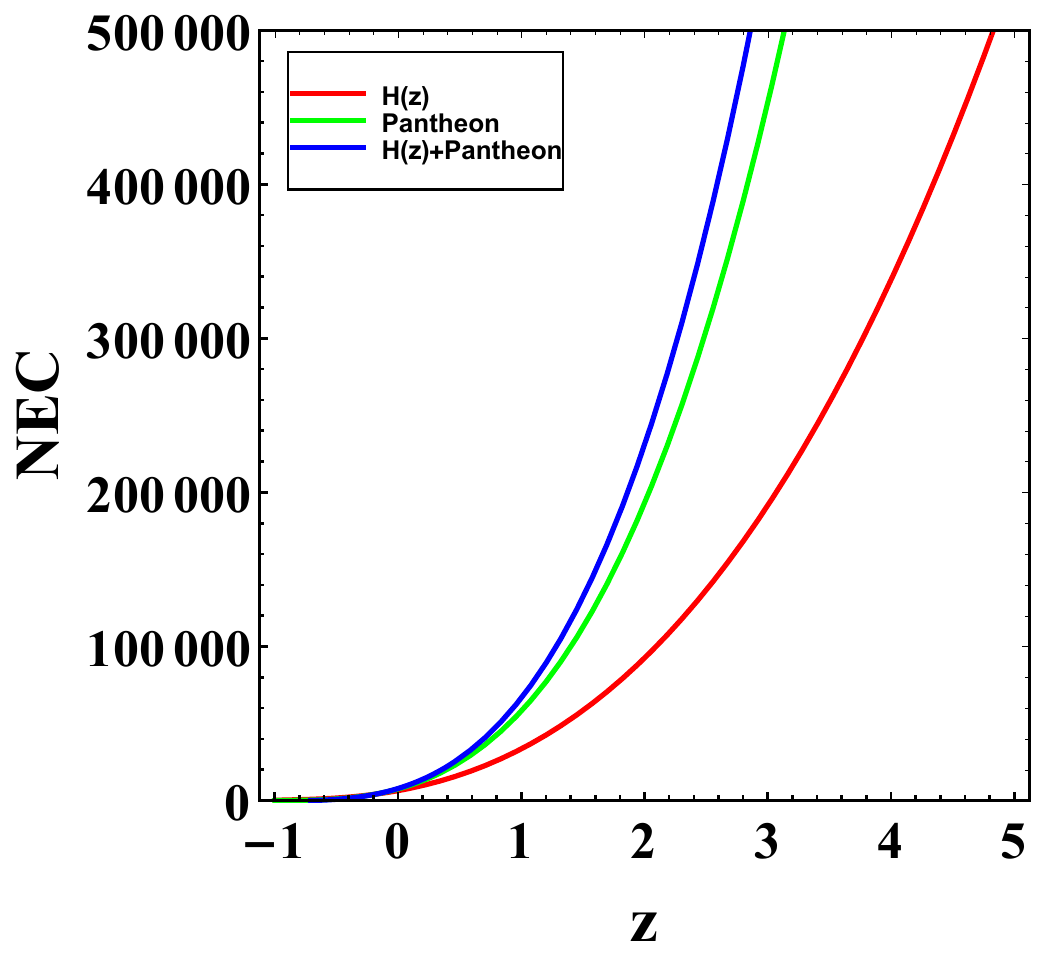}}\hfill
	\subfloat[]{\label{b}\includegraphics[scale=0.43]{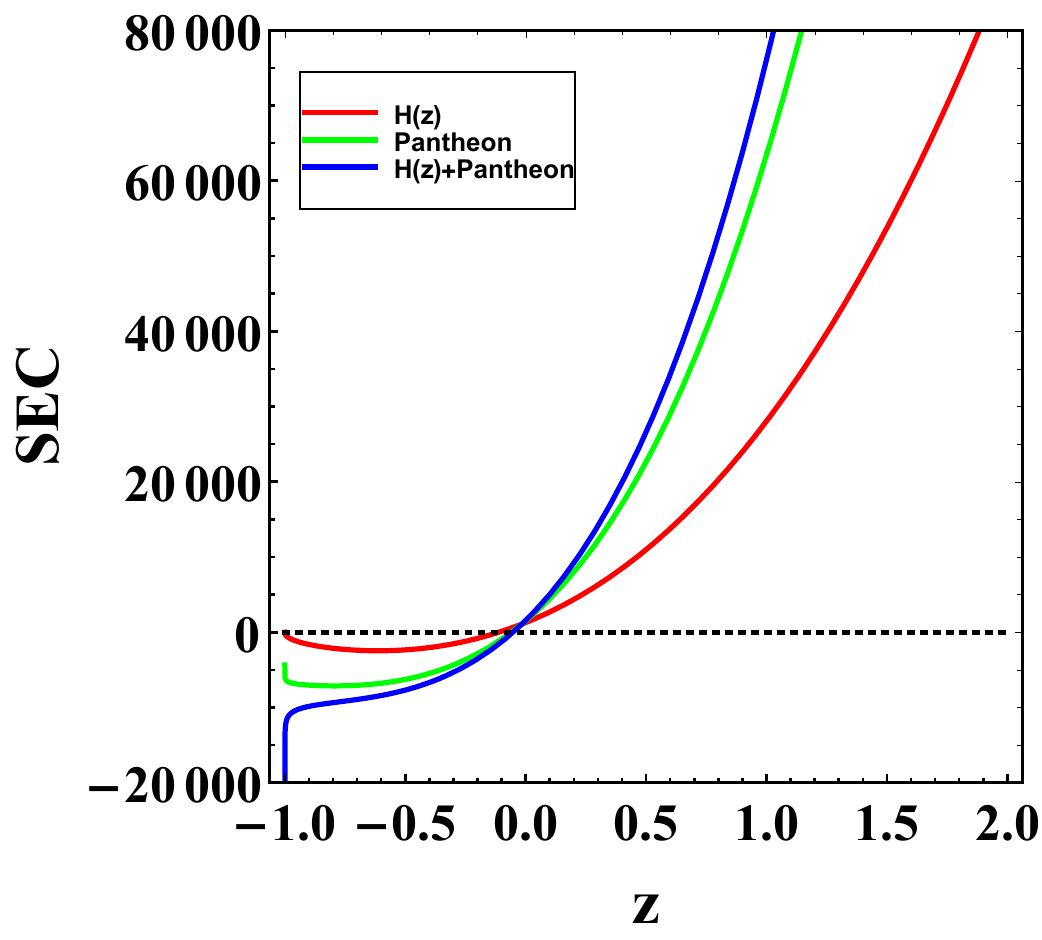}}\par 
	\subfloat[]{\label{c}\includegraphics[scale=0.43]{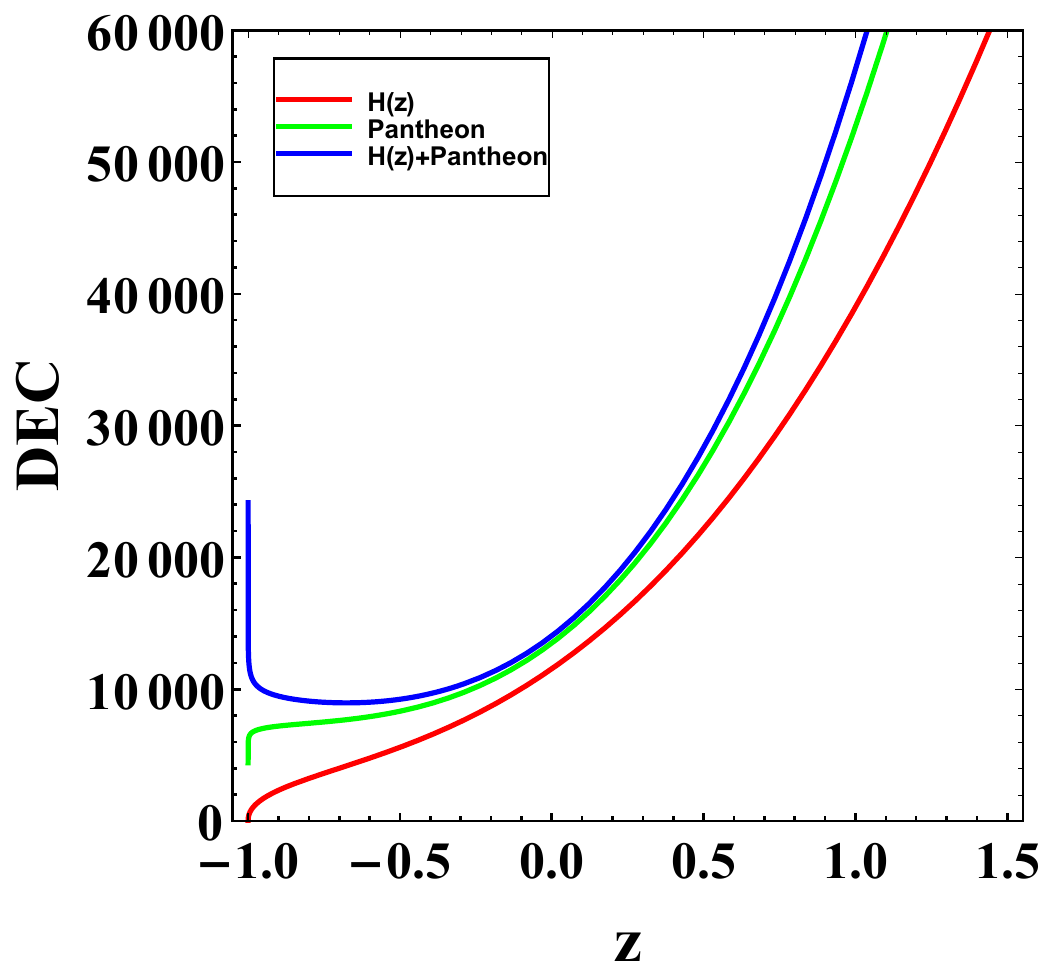}}
	\caption{ The plots for NEC, SEC, and DEC for the mentioned observational datasets.}
 \label{ecs}
\end{figure}

In Fig. \ref{ecs}, our probe reveals that for all observational datasets, NEC and DEC hold for the whole range of $ z $, whereas SEC violates in the future.    

\section{Dynamical System Stability} \label{dss}
In this section, we employ the dynamical system technique to assess the stability of the system. To achieve this, we initially select a set of dimensionless variables as follows
\begin{equation}\label{26}
	\Omega_M = \frac{\kappa \rho}{3 H^2}, \quad  y = \frac{\kappa p}{H^2}, \quad \delta = \kappa \lambda.
\end{equation}
The variable $\Omega_M$ represents a time-dependent quantity, signifying the fractional energy density budget of the universe, while $y$ captures the dynamics of the corresponding pressure. We introduce a dimensionless constant, denoted as $\delta$, which can take any non-zero value. Utilizing Eq.~\eqref{17} and Eq.~\eqref{18}, we express
\begin{equation}\label{27}
	H^2 = \frac{\kappa (3 p \kappa \lambda -\rho + 3 \kappa \lambda \rho )}{-3 + 12 \kappa \lambda},
\end{equation}
and 
\begin{equation}\label{28}
	\dot{H} = \frac{-\kappa}{2} (\rho + p).
\end{equation}
Expressing Eq.~\eqref{27} in terms of the dynamical variables that constrain $y$, we can represent this constraint in terms of $\om$ as 
\begin{equation}\label{29}
	y = \frac{-1+\om}{\delta} + 4-3 \om.
\end{equation}
To analyze the dynamics of the system, we construct the autonomous system by calculating the first-order time derivative of the dynamical variable, as follows 
\begin{equation}\label{30}
	\om' = \frac{d \om}{H dt} = -(1-4 \delta)(y+3\om) - 2 \delta \left(j+ \frac{3}{2}(y+3\om) - 1\right) + \om(y+3\om).
\end{equation}
Here, $H dt = dN = d \log a$. We have used $j = \frac{\dddot{a}}{a H^3}$ as a constant parameter. This constraint allows the evaluation of the system's dynamics using a single parameter, resulting in a 1D phase space. The equation of state and deceleration parameter can then be expressed in terms of dynamical variables as
\begin{equation}\label{31}
	\omega = \frac{p}{\rho} = \frac{y}{3\om},  \quad q = -1 + \frac{1}{2}(y+3\om).
\end{equation}
We impose an additional constraint on the fractional energy density parameter, necessitating it to remain positive and bounded between $0$ and $1$,
\begin{equation}\label{32}
	0 < \Omega_M \le 1.
\end{equation} 
The critical points that satisfy this constraint are the only ones that are considered physically viable. The system produces two critical points, which are tabulated in Table-\ref{tab:critical_pts}. The qualitative behavior of these points is discussed in detail.  
\begin{table}[t]
	\centering
		\caption{The critical points and their nature}
	\begin{tabular}{c c c c }
		\hline
		\hline
		Points & $\Omega_{M*}$ & $\omega$  & $d\Omega_{M}'/d\om \vert_{\Omega_{M*}}$\\
		\hline
		$P_{1}$ & $1-\frac{1}{2} \delta  \left(\sqrt{8 j+1}+5\right)$ & Fig. \ref{fig:eos_p1}  & $-\sqrt{8 j+1}$\\
		\hline
		$P_{2}$ & $\frac{1}{2} \delta  \left(\sqrt{8 j+1}-5\right)+1$ & Fig. \ref{fig:eos_p2}  & $\sqrt{8 j+1}$\\
		\hline
		\hline
	\end{tabular}

	\label{tab:critical_pts}
\end{table}
\begin{figure}\centering
	\subfloat[\label{fig:x_existence_p1}]{\includegraphics[scale=0.50]{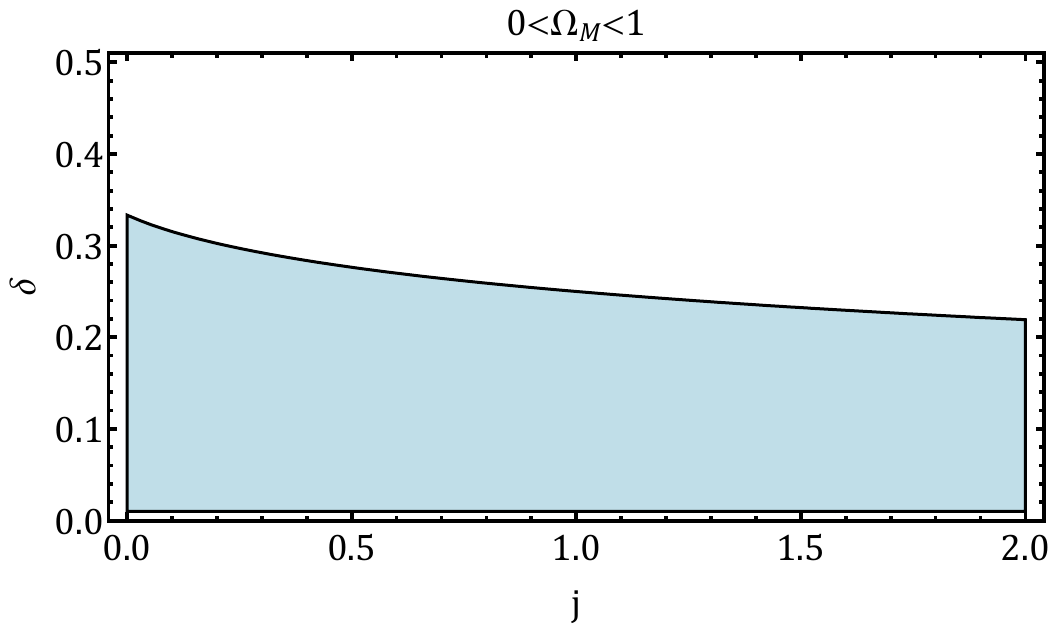}}
	\subfloat[\label{fig:eos_p1}]{\includegraphics[scale=0.5]{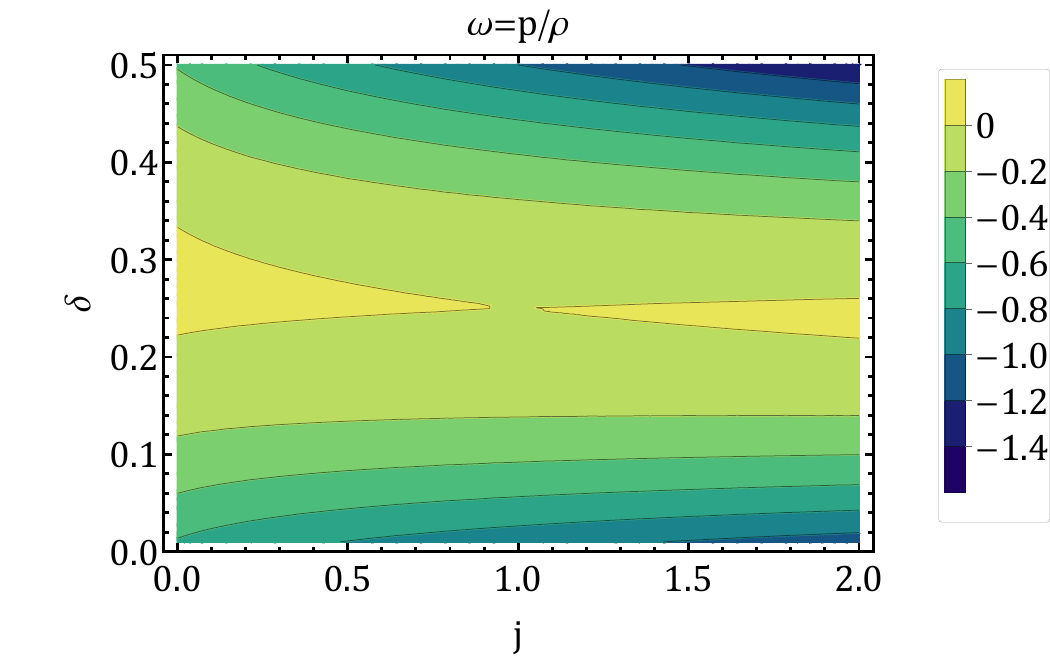}}
	\caption{ The plots for the constrained relation Eq.~\eqref{32} and EoS $\omega$ in the parametric space of $(j,\delta)$ for the fixed point $P_{1}$.}
\end{figure}
\begin{itemize}
	\item \textbf{Point $P_{1}$:} This critical point is dependent on both $j$ and $\delta$ and becomes a real point for $j > -1/8$. However, it cannot assume arbitrary values and must adhere to the constraint given by Eq.~\eqref{32}. The corresponding region of its existence has been plotted in Fig. \ref{fig:x_existence_p1}. The nature of the fixed point can be inferred from the equation of state $\omega$, as illustrated in Fig. \ref{fig:eos_p1}, further constraining the values of $j$ and $\delta$. The contour plot reveals that the point can exhibit both accelerating and non-accelerating solutions for $\delta < 0.3$. However, for $\delta \ge 0.5$, the point yields a phantom solution; nevertheless, the fractional energy density violates the constraint relation. To assess the stability of the fixed point, we determined the first derivative $\frac{d\Omega_{M}'}{d\om} \Big|_{\Omega_{M*}}$ at this point, resulting in a negative quantity. Following the stability discussion presented in Appendix \ref{ap}, we conclude that as $N \to +\infty$, the point becomes a stable fixed point corresponding to the future epoch of the universe. Conversely, at the past epoch, i.e., $N \to - \infty$, the point becomes unstable.
	\begin{figure}\centering
		\subfloat[\label{fig:x_existence_p2}]{\includegraphics[scale=0.50]{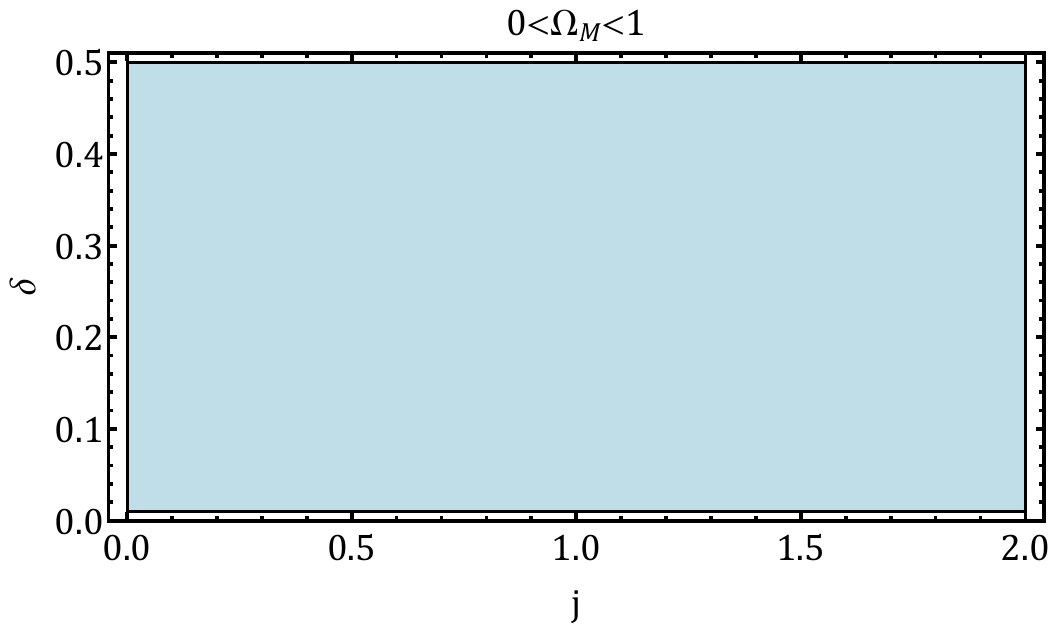}}
		\subfloat[\label{fig:eos_p2}]{\includegraphics[scale=0.5]{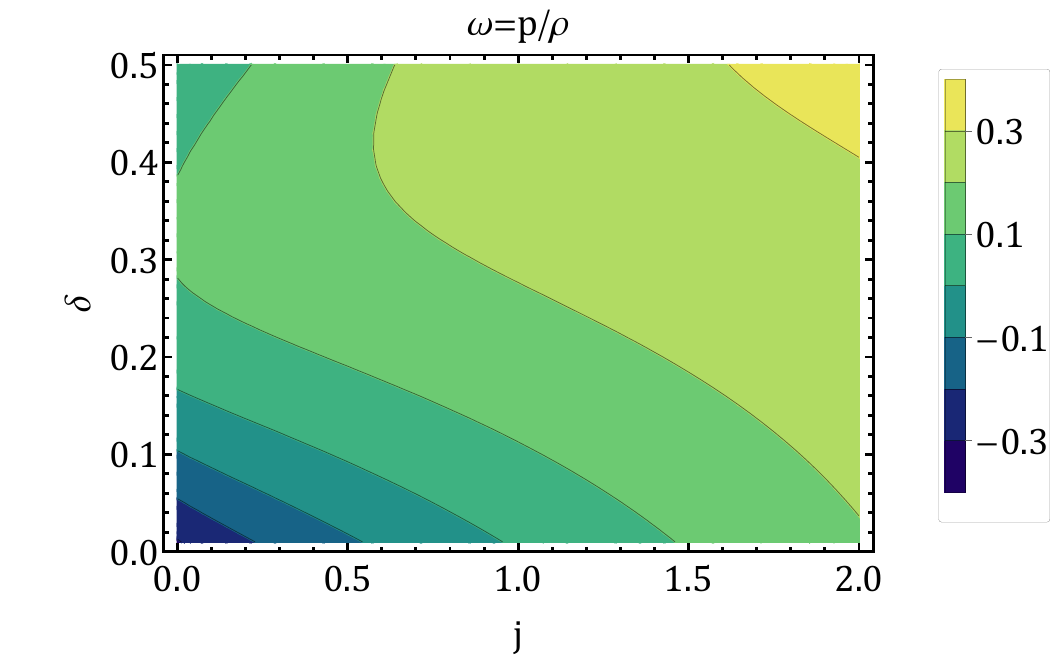}}
		\caption{ The plots for the constrained relation Eq.~\eqref{32} and EoS parameter $\omega$ in the parametric space of $(j,\delta)$ for the fixed point $P_{2}$.}
	\end{figure}
	
	\begin{figure}[t]\centering
		\subfloat{\includegraphics[scale=0.5]{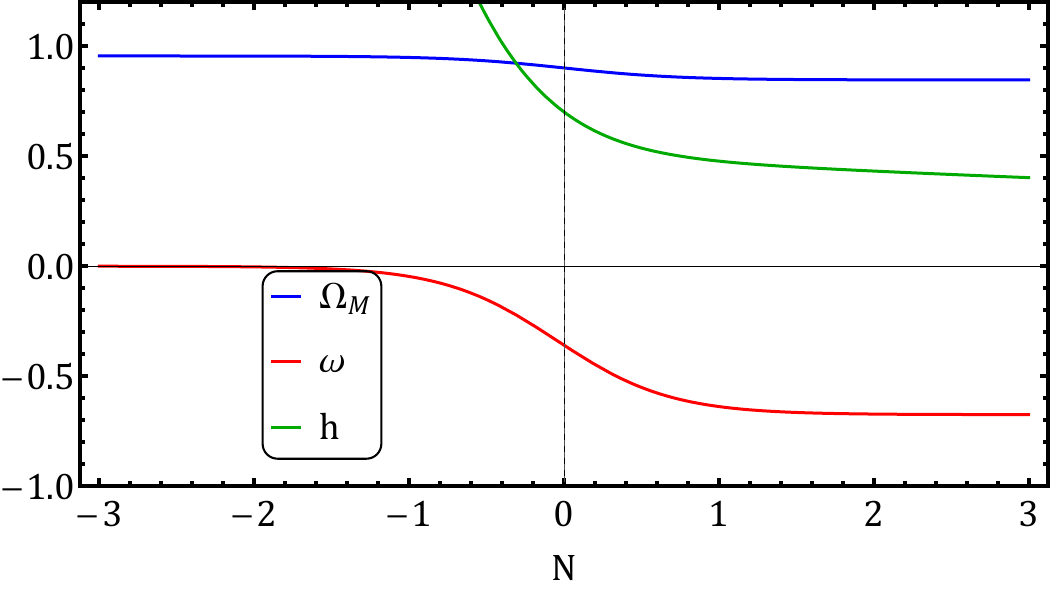}}
		\caption{The evolution of cosmological parameters for $\delta = 0.04, 	\ j = 0.8$.}
		\label{fig:evo_1}
	\end{figure}

	\item \textbf{Point $P_{2}$:} The constraint of the point has been visualized in Fig. \ref{fig:x_existence_p2}, illustrating its existence for $j > -1/8$. The equation of state corresponding to this point is depicted in Fig. \ref{fig:eos_p2}, showing positive values and indicating non-accelerating characteristics. For $j > 1.5$ and $\delta > 0.4$, the point can mimic the radiation equation of state. However, within this range, the aforementioned point becomes unphysical. In the range $0.5 < j < 1.5$ and $0.01 < \delta < 0.2$, this point can exhibit characteristics of a pressure-less fluid. Upon evaluating stability, the derivative $\frac{d\Omega_{M}'}{d\om} \Big|_{\Omega_{M*}}$ yields a positive value, indicating stability in the past epoch.
\end{itemize}

From the analysis of critical points and based on their characteristic properties, we have constrained $j$ and $\delta$. One of the critical points is found to be stable, producing an accelerating solution that represents the present epoch of the universe. To further analyze the system's dynamics, we evaluate the evolution of cosmological parameters, i.e., $(\Omega_M, \omega, h \equiv H/100)$ against $N = \log a$ in Fig. \ref{fig:evo_1}, spanning from the past epoch to the future epoch. We select benchmark points $(j=0.8, \delta = 0.04)$ to ensure the existence of both fixed points simultaneously. In the past epoch, the fractional energy density dominates, and the corresponding equation of state is near zero, indicating a matter-dominated phase of the universe. As the system evolves toward the present epoch, the equation of state increases toward negative values. In the future epoch, $\omega$ saturates to $-0.70$, and energy density saturates to $0.90$. Hence, the system exhibits quintessence-like behavior in the late-time epoch and produces matter characteristics in the past epoch of the universe. We have also depicted the evolution of the Hubble parameter $h(N) = H(N)/100$ using
\begin{equation}\label{33}
	\frac{dh}{dN} = -\frac{3}{2}h(N) \left(\frac{y}{3}+x\right).
\end{equation}
The behavior indicates that in the past, the Hubble parameter increases, and as the system progresses into the future epoch, the Hubble parameter saturates to lower values.

\section{Conclusions}\label{con}
\quad In this work, we study Rastall's theory of gravity, which is based on a variational principle. In this theory, we consider a non-minimal coupling between geometry and matter fields and discuss the recent theoretical developments in alternative theories of general relativity. This work is framed for the flat, isotropic, and homogeneous universe in the FLRW metric, where we have obtained the EFEs with the calculated value of the Hubble parameter in terms of the jerk parameter. Furthermore, we examine the essential physical and dynamical parameters one by one. We evaluate the best-fitted values of model parameters $ \alpha $ and $ j $ by the MCMC method, where we use the Hubble dataset, the $ Pantheon $ dataset, and their joint dataset.

The trajectories of the deceleration parameter for these best-fit values indicate the phase transition from deceleration to acceleration. At present, our model shows an accelerating universe.

Also, if we look at the behavior of energy density, it has monotonically decreased from high redshift to low redshift for all observational datasets. The isotropic pressure $ p $ is positive in the early universe and holds a negative value at present as well as in the future, which indicates the presence of repulsive force and an accelerating expansion of the universe. According to the results of the EoS parameter, our model is a quintessence dark energy model at present. In Fig. \ref{rhopw}, $ \omega $ depicts the different dominating eras of the universe.

In this model, NEC and DEC both hold good, but SEC violates, which shows the existence of exotic matter in the universe. Hence, finally, we conclude that our model, which is studied in Rastall gravity, is a quintessence dark energy model.  

Finally, we conducted a dynamical stability analysis of the system, treating the jerk parameter $j$ as a constant quantity, which imposes a vital constraint on the system. Our findings revealed a 1D phase space, with the system yielding two critical points. One critical point results in a stable, accelerating solution in the future epoch, resembling the behavior of a quintessential dark energy model. The other critical point exhibits non-accelerating characteristics, similar to dark matter fluid, and stabilizes at the past epoch. Thus, the qualitative behavior of the system demonstrates that Rastall's theory of gravity can consistently account for different phases, namely the matter-dominated and accelerating epochs of the universe.

\vskip0.3in

\textbf{\noindent Acknowledgements}\\
 J. K. Singh wishes to thank Prof. Joao Luis Rosa, Institute of Physics, University of Tartu, W. Ostwaldi 1, 50411 Tartu, Estonia, for fruitful discussions. 

\vskip0.2in

\textbf{\noindent Data Availability Statement}\\
No new data were created or analyzed in this study.

\vskip0.2in

\textbf{\noindent Code Availability Statement}\\
No new code was created in this study.

\appendix

\section{Dynamical system in 1D} \label{ap}

Consider the autonomous equation in 1D as,  
\begin{equation}\label{app1}
	\dfrac{dx}{dN} =\dot{x} = f(x).
\end{equation}
Suppose $x_{0} \in \mathbb{R}^n$ is a fixed point of the autonomous equation such that $f(x_0) = 0$. To determine the stability of the system at this fixed point, we perturb the system around it as,
\begin{equation}\label{app2}
	x = x_0 + \epsilon,
\end{equation}
where, $\epsilon\ll 1$. We expand $f(x)$ using Taylor's series around the fixed point $x_0$ as, 
\begin{equation}\label{app3}
	f(x) = f(x_0) + \dfrac{df}{dx}\Big|_{x_0} (x-x_0) + ...
\end{equation}
Inserting it into the autonomous equation Eq.~\eqref{app1} and truncating the series after the first order, we obtain,
\begin{equation}\label{app4}
	\begin{split}
		\dot{\epsilon} & = f'(x_0) \epsilon ,\\
		\implies 	\epsilon & = C \exp(f'(x_0) \, N).
	\end{split}
\end{equation}
Here, $C$ is an integration constant. Observing Eq.~\eqref{app4}, as $N \rightarrow + \infty$, the perturbation increases (decreases) when $f'(x_0) > 0 \ (<0)$, indicating the system becomes unstable (stable) at that point.

\end{document}